\documentclass[conference]{IEEEtran}
\IEEEoverridecommandlockouts
\usepackage{amsmath, amssymb, amsfonts, bm, mathtools}
\usepackage{amsthm}

\usepackage{graphicx}
\usepackage{subfigure}
\usepackage{tikz}
\usetikzlibrary{
  arrows.meta, 
  decorations.markings, 
  circuits, 
  automata,
  arrows, 
  positioning, 
  calc
}

\usepackage{booktabs}
\usepackage{makecell}

\usepackage{textcomp}
\usepackage{verbatim}
\usepackage{xcolor}

\usepackage{cite}
\usepackage[hidelinks,bookmarks=false]{hyperref}
\usepackage[capitalise,noabbrev]{cleveref}
\Crefformat{figure}{#2Fig.~#1#3}
\Crefformat{equation}{#2Eq.~(#1)#3}
\crefname{equation}{Eq.}{Eqs.}
\crefname{figure}{Fig.}{Figs.}

\newcommand{\ind}{\mathbf{1}}

\newcommand{\draftgraphic}[3]{%
  \IfFileExists{#3}{%
    \includegraphics[#1]{#3}%
  }{%
    \fbox{\parbox[c][#2][c]{0.94\linewidth}{\centering\scriptsize
    Corrected figure pending. Re-run the corrected simulator.}}%
  }%
}

\usepackage{titlesec}
\titlespacing\section{0pt}{6pt plus 2pt minus 2pt}{4pt plus 2pt minus 2pt}
\titlespacing\subsection{0pt}{4pt plus 2pt minus 2pt}{3pt plus 2pt minus 2pt}

\begin{document}
\bstctlcite{IEEEexample:BSTcontrol}

\title{MAQR: Lightweight Memory-Aware Q-Routing for Fresh Status Delivery in Multi-Hop IoBNTs}

\newif\ifanonymous
\anonymousfalse    

\ifanonymous

\author{
    \IEEEauthorblockN{Anonymous Authors}
}

\else

\author{
    \IEEEauthorblockN{
        Ruifeng Zheng\IEEEauthorrefmark{1},
        Veronika Volkova\IEEEauthorrefmark{1}, 
        Zhihan Xu\IEEEauthorrefmark{1}, 
        Pengjie Zhou\IEEEauthorrefmark{1}\IEEEauthorrefmark{3},
        Martín Schottlender\IEEEauthorrefmark{1},\\
        Rafael F. Schaefer\IEEEauthorrefmark{2}\IEEEauthorrefmark{3},
        Frank H.\,P. Fitzek\IEEEauthorrefmark{1}\IEEEauthorrefmark{3}, and 
        Pit Hofmann\IEEEauthorrefmark{1}\IEEEauthorrefmark{3}\\[1ex]
    }
    \IEEEauthorblockA{
         \IEEEauthorrefmark{1}Deutsche Telekom Chair of Communication Networks, Dresden University of Technology, Germany\\
         \IEEEauthorrefmark{2}Chair of Information Theory and Machine Learning, Dresden University of Technology, Germany\\
         \IEEEauthorrefmark{3}Centre for Tactile Internet with Human-in-the-Loop (CeTI), Dresden, Germany\\[0.5ex]
         \texttt{\{firstname.lastname\}@tu-dresden.de}
     }
    \thanks{This work was supported by the German Research Foundation (DFG) as part of Germany's Excellence Strategy -- EXC 2050/2 -- Cluster of Excellence ``Centre for Tactile Internet with Human-in-the-Loop'' (CeTI) of TUD Dresden University of Technology under project ID 390696704, and by the German Federal Ministry of Research, Technology and Space of Germany (BMFTR) as part of the research program Communication Systems ``Souverän. Digital. Vernetzt.''. Joint project 6G-life, project id number 16KIS2413K, the program “Verbundprojekt: Disruptive Kommunikationsparadigmen für technologische Souveränität, Resilienz und Shared Prosperity -- Translation in Industrie und Aufbau innovativer Technologiedemonstratoren -- CommUnity,” project ID number 16KISS012K, and the project ``Internet of Bio-Nano-Things'' (IoBNT) with project id number 16KIS1994.}
}

\fi

\maketitle

\begin{abstract}
This paper proposes memory-aware Q-routing (MAQR) for fresh status delivery over history-dependent molecular links in multi-hop Internet of Bio-Nano Things (IoBNT) networks. Repeated transmissions leave residual molecules, making link reliability history-dependent and fixed routing inefficient. Each link is calibrated from a spatial Markov model and approximated by three nonnegative decay modes. The resulting reduced memory state and packet bit composition provide an instantaneous link-success estimate. MAQR combines quantized memory and reliability descriptors with packet age, queue occupancy, and molecule reserve in a lightweight partially observable Q-routing heuristic. It is trained offline and requires only reduced-state updates and Q-table lookup during deployment. Across five independent training pairs and 20 shared evaluation seeds, MAQR reduces mean AoI by $46.9\%$ relative to memory-blind Q-routing at $0.30$ packets/slot while improving molecular efficiency. Frozen-policy sweeps further show comparable performance under weak memory and an increasing MAQR advantage as channel memory strengthens.
\end{abstract}

\begin{IEEEkeywords}
Age of information, Internet of Bio-Nano Things, molecular communication, molecular channel memory, Q-routing, reinforcement learning.
\end{IEEEkeywords}

\section{Introduction}

\IEEEPARstart{T}{he} Internet of Bio-Nano Things (IoBNT) envisions networks of bio-nano sensors, engineered cells, and bio-cyber interfaces for in-body sensing and actuation~\cite{akyildiz2015internet,lee2023internet}. Molecular communication (MC) provides a biocompatible means of exchanging information at such small scales~\cite{kuscu2019transmitter}. Since the range of an individual molecular link is limited, decode-and-forward relays can extend communication coverage and enable multi-hop delivery of physiological status updates to a bio-cyber gateway~\cite{einolghozati2013relaying,ahmadzadeh2015analysis}. One biologically inspired example is bacterial nanonetworking, in which natural properties such as motility, chemotaxis, and conjugation can be exploited to transfer information through intermediate bacterial populations~\cite{balasubramaniam2013multi,balasubramaniam2014exploiting}.

A key challenge in such networks is that molecular-link reliability depends on recent transmission history. Diffusion, advection, and reversible reception produce long channel tails, leaving residual molecules that affect subsequent observations~\cite{zheng2025anis}. Repeated use of the same link can therefore degrade its instantaneous reliability, making a fixed shortest or average-reliability route inefficient. The resulting decoding failures and retransmissions increase molecule consumption and delay fresh status delivery. This is particularly important for monitoring applications, where the age of information (AoI) captures the freshness of the latest update available at the destination~\cite{sun2017update}. Point-to-point molecular channels already exhibit a direct tradeoff between update frequency, intersymbol interference, and AoI~\cite{torres2022age}.

In broader communication networks, recent studies have investigated AoI optimization in multichannel multi-hop wireless systems, freshness-aware next-hop selection in low-power IoT infrastructures, and learning-based routing under time-varying topologies~\cite{lou2022age,gao2025toward}. Collectively, these studies highlight the importance of adapting forwarding decisions to information freshness and evolving network conditions. Existing multi-hop MC studies, however, have largely focused on relay placement, hop count, and routing based on average link quality or energy expenditure~\cite{cheng2020energy,cheng2022joint,ansari2023multi}. Opportunistic molecular routing has also used distance and concentration-gradient information to select relays~\cite{aijaz2015opportunistic}. These approaches do not explicitly adapt routing decisions to the instantaneous transmission-induced memory state of multiple outgoing molecular links. Directly exposing the full physical channel state to the routing layer would also introduce excessive complexity for resource-constrained IoBNT nodes. A compact cross-layer representation of molecular memory is therefore needed to support history-aware routing.

To address this problem, we propose \emph{memory-aware Q-routing} (MAQR) for freshness-aware packet forwarding in multi-hop IoBNT networks. MAQR uses a reduced representation of residual molecular memory and link reliability within a lightweight tabular Q-routing framework, with training performed offline to avoid online exploration. Our main contributions are as follows:
\begin{itemize}
\item We formulate fresh status delivery over history-dependent molecular links using a consistent single-bit response, packet-dependent molecular release, and composition-conditioned packet reliability.
\item We reduce each spatial Markov channel to three positive decay modes and expose only quantized residual-memory and reliability descriptors to the routing layer.
\item We develop an offline-trained tabular MAQR heuristic and compare it with a state-matched memory-blind agent and a dynamic non-learning memory-aware baseline.
\end{itemize}

\textit{Notation:} Italic, bold lowercase, bold uppercase, and calligraphic letters denote scalars, column vectors, matrices, and sets, respectively. The symbols $|\mathcal{A}|$, $(\cdot)^{\mathsf T}$, and $\ind\{\cdot\}$ denote cardinality, transpose, and the indicator function, respectively; $[x]^+\triangleq\max\{x,0\}$.

\section{System Model}
\label{sec:system_model}

\begin{figure}[t]
\centering
\usetikzlibrary{arrows.meta,positioning,fit}

\begin{tikzpicture}[
    >={Latex[length=1.8mm]},
    font=\scriptsize,
    nd/.style={draw,circle,minimum size=6.5mm,inner sep=0pt,fill=white},
    box/.style={draw,rounded corners=2pt,align=center,text width=11mm,
                minimum height=7mm,inner sep=1.5pt,fill=white},
    cand/.style={->,densely dashed,gray!60!black},
    sel/.style={->,very thick,red!70!black},
    cyber/.style={->,thick,densely dashdotted,blue!70!black},
    lbl/.style={font=\scriptsize,inner sep=1pt}
]
\node[nd,fill=gray!15] (S)  at (0,1.9)   {$s$};
\node[nd]              (R1) at (1.6,2.8) {$r_1$};
\node[nd]              (R2) at (1.6,1.0) {$r_2$};
\node[nd]              (R3) at (3.4,1.9) {$r_3$};
\node[nd]              (R4) at (3.4,0.5) {$r_4$};
\node[nd,fill=gray!30] (G)  at (5.2,1.9) {$g$};
\node[box,fill=gray!8] (M)  at (6.9,1.9) {medical\\server};

\draw[cand] (S)  -- (R1);
\draw[cand] (S)  -- (R2);
\draw[cand] (R1) -- (R3);
\draw[cand] (R2) -- (R3);
\draw[cand] (R2) -- (R4);
\draw[cand] (R3) -- (G);
\draw[cand] (R4) -- (G);
\draw[sel]  (R1) to[bend left=0]
    node[midway,above,sloped,lbl,red!70!black]{max residual memory} (G);
\draw[cyber] (G) -- node[midway,above,lbl,blue!70!black]{cyber} (M);

\node[box] (B1) at (0.15,-1.1) {receive};
\node[box,right=3.5mm of B1] (B2) {decode\\\& verify};
\node[box,right=3.5mm of B2] (B3) {store};
\node[box,right=3.5mm of B3,fill=gray!15] (B4) {select\\next hop};
\node[box,right=3.5mm of B4] (B5) {re-encode\\\& release};
\foreach \a/\b in {B1/B2,B2/B3,B3/B4,B4/B5}{\draw[->] (\a) -- (\b);}
\node[lbl,anchor=west] at (0.05,-0.35)
    {Decode-and-forward operation at router $r_i$};

\begin{scope}[shift={(0,-2.15)}]
  \draw[cand] (0.15,0) -- ++(0.6,0) node[right,lbl]{candidate link};
  \draw[sel]  (2.5,0)  -- ++(0.6,0) node[right,lbl]{selected route};
  \draw[cyber](4.9,0)  -- ++(0.6,0) node[right,lbl]{cyber link};
\end{scope}
\end{tikzpicture}
\caption{Multi-hop IoBNT architecture and decode-and-forward operation. A shorter route may be temporarily worse because repeated activation leaves residual molecules. Link parameters are given as $(d_{ij},v_{ij})$, with distance in $\mu\mathrm{m}$ and flow velocity in $\mu\mathrm{m}/\mathrm{s}$: $s\!\to\!r_1$: $(18,6)$; $s\!\to\!r_2$: $(14,16)$; $r_1\!\to\!r_3$: $(15,6)$; $r_2\!\to\!r_3$: $(24,16)$; $r_2\!\to\!r_4$: $(16,16)$; $r_1\!\to\!g$: $(29,6)$; $r_3\!\to\!g$: $(14,16)$; and $r_4\!\to\!g$: $(18,16)$.}
\label{fig:system_model}
\vspace{-0.5cm}
\end{figure}
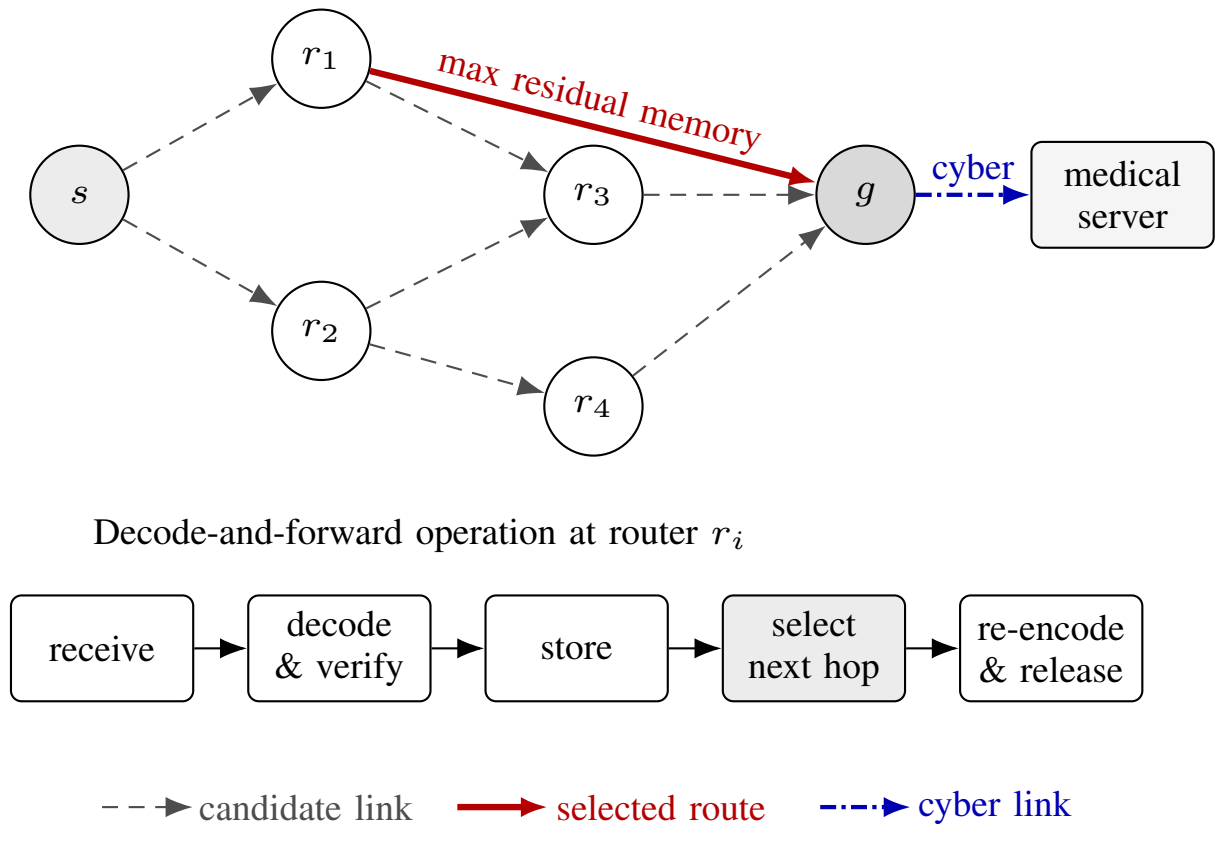

\subsection{Network, Packet, and Resource Model}
\label{subsec:network_packet_model}

As illustrated in~\cref{fig:system_model}, we consider a multi-hop IoBNT represented by a directed graph $\mathcal{G}=(\mathcal{V},\mathcal{E})$ with $\mathcal{V}=\{s\}\cup\mathcal{V}_{\mathrm{R}}\cup\{g\}$, where $s$ is a bio-nano sensing node, $\mathcal{V}_{\mathrm{R}}$ is the set of decode-and-forward molecular routers, and $g$ is the bio-cyber gateway. A directed edge $(i,j)\in\mathcal{E}$ denotes a feasible molecular link, and $\mathcal{N}_i^{+}=\{j:(i,j)\in\mathcal{E}\}$ denotes the outgoing neighbor set of node $i$. Each link $(i,j)$ is characterized by its propagation distance $d_{ij}$ and flow velocity $v_{ij}$, which determine its propagation and memory characteristics together with the remaining physical channel parameters.

We consider a time-slotted network, where each network slot has duration $T_s$.
Let $A[t]\in\{0,1\}$ denote the update-arrival indicator at the sensing node $s$ in slot $t$, where $A[t]=1$ indicates that a new status update is generated at the beginning of slot $t$, and $A[t]=0$ otherwise. The update arrivals are modeled as an i.i.d. Bernoulli process~\cite{kadota2021minimizing}, i.e.,
\begin{equation}
\Pr\{A[t]=1\}=\lambda,\qquad
\Pr\{A[t]=0\}=1-\lambda,
\label{eq:packet_arrival}
\end{equation}
where $\lambda=\mathbb{E}[A[t]]$ denotes the average update-generation rate in packets per slot. The sensing node employs freshest-update replacement: whenever a new update is generated, any undelivered packet waiting at the source is replaced by the newly generated packet. Intermediate routers employ first-in first-out (FIFO) queues with finite capacity $Q_{\max}$.

Each packet $p$ is represented by the binary sequence
\begin{equation}
    \bm{\beta}_p=
    \begin{bmatrix}
    \beta_p[0],\ldots,\beta_p[B_{\mathrm{pkt}}-1]
    \end{bmatrix}^{\top},
    \qquad
    \beta_p[q]\in\{0,1\},
    \label{eq:packet_bit_sequence}
\end{equation}
where $B_{\mathrm{pkt}}$ denotes the packet length in bits. The packet is transmitted within one network slot using $B_{\mathrm{pkt}}$ equal bit intervals of duration $T_b=T_s/B_{\mathrm{pkt}}$. Bit $q$ is released at the beginning of its interval and sampled after a fixed link decision delay $\tau_{ij}^{\mathrm{dec}}<T_b$. The packet bits are conveyed over each molecular link using binary concentration-shift keying (BCSK), where a bit-1 is represented by the release of 
$N_{\mathrm{tx}}$ signaling molecules and a bit-0 by no release. To characterize the resulting molecular cost, we define
\begin{equation}
    B_p^{(1)}=\sum_{q=0}^{B_{\mathrm{pkt}}-1}\beta_p[q]
    \label{eq:number_of_one_bits}
\end{equation}
as the number of bit-1 symbols in packet $p$. Accordingly, one transmission attempt consumes a number of
\begin{equation}
    M_p^{\mathrm{tx}}=B_p^{(1)}N_{\mathrm{tx}}
    \label{eq:packet_molecule_cost}
\end{equation}
signaling molecules. Each packet transmission attempt occupies one network slot. Packet errors are assumed to be perfectly detected, e.g., by an ideal cyclic redundancy check (CRC), whose signaling overhead is not explicitly modeled.

At the network layer, packet forwarding is controlled through link-activation decisions. Let $\sigma_{ij}[t]\in\{0,1\}$ indicate whether node $i$ activates link $(i,j)$ in slot $t$. Each node activates at most one outgoing link per slot, i.e.,
\begin{equation}
    \sum_{j\in\mathcal{N}_i^{+}}\sigma_{ij}[t]\leq 1.
    \label{eq:single_link_activation}
\end{equation}
When $\sigma_{ij}[t]=1$, node $i$ attempts to transmit its head-of-line packet to node $j$. If the packet is successfully decoded and no packet error is detected, it is removed from the transmitter queue and inserted into the receiver queue; if $j=g$, the packet is considered delivered and leaves the network. Otherwise, the packet remains at node $i$ for a later retransmission. All routing decisions in slot $t$ are based on the queue states observed at the beginning of that slot. Successful arrivals are appended after all beginning-of-slot decisions have been resolved, and a packet forwarded to an intermediate router becomes eligible for transmission only in the following slot. If simultaneous arrivals exceed the remaining buffer capacity, they are ordered by decreasing generation time and then by sequence identifier; excess packets are dropped.

In addition to link activation, packet transmission is constrained by the signaling-molecule reserve available at each transmitting node. Let $E_i[t]$ denote the reserve of node $i$ before replenishment in slot $t$, $E_{\max}$ the reserve capacity, and $E_{\mathrm{rec}}$ the number of molecules replenished per slot. The reserve available for transmission is
\begin{equation}
    \widetilde E_i[t]=\min\left\{E_{\max},E_i[t]+E_{\mathrm{rec}}\right\}.
    \label{eq:reserve_recovery}
\end{equation}
A packet $p$ can be transmitted only if $\widetilde E_i[t]\geq M_p^{\mathrm{tx}}$. Following the transmission decision, the reserve evolves as
\begin{equation}
    E_i[t+1]=\widetilde E_i[t]-M_p^{\mathrm{tx}}\sum_{j\in\mathcal{N}_i^{+}}\sigma_{ij}[t],
    \label{eq:reserve_update}
\end{equation}
where the summation is either zero or one due to~\cref{eq:single_link_activation}. Thus, a transmission attempt consumes $M_p^{\mathrm{tx}}$ molecules regardless of the decoding outcome, whereas an idle node retains its replenished reserve.

Throughout this work, hop-by-hop acknowledgments are assumed to be error-free, with negligible transmission duration. Potentially interfering molecular links are orthogonalized through distinguishable molecule types or nonoverlapping transmission schedules. Consequently, inter-link interference 
is neglected, and the residual state of each directed link depends only on its own transmission history.

\subsection{Physical Link and Reduced Memory Model}
\label{subsec:reduced_link_model}

Each directed molecular link $(i,j)\in\mathcal{E}$ is first characterized using the spatial transient Markov model in~\cite{zheng2025DNA-Based,zheng2026system,zheng2026markov}. At physical time step $k$, the transient molecular state evolves as
\begin{equation}
    \bm{x}_{ij}[k+1]=\bm{P}_{ij}\left(\bm{x}_{ij}[k]+\bm{b}_{ij}u_{ij}[k]\right),
    \label{eq:physical_link_state}
\end{equation}
where $\bm{x}_{ij}[k]$ contains the expected molecular populations in the transient spatial and receiver states, $\bm{P}_{ij}$ is the transient transition matrix, $\bm{b}_{ij}$ specifies the injection state, and $u_{ij}[k]$ denotes the molecular input. The transition matrix $\bm{P}_{ij}$ is determined by the diffusion coefficient, link geometry, fixed flow velocity, reversible binding kinetics, and downstream flow-out. This full spatial model is used only for offline link characterization and is not propagated by the routing algorithm.

We define $\bm{o}_{ij}$ as the vector selecting the receiver-observation state, and $\bm{A}_{ij}^{\mathrm{dec}}$ and $\bm{A}_{ij}^{\mathrm{slot}}$ as the propagation matrices over $\tau_{ij}^{\mathrm{dec}}$ and one network slot, respectively. On this common physical scale, the sampled response to a bit-1 release of $N_{\mathrm{tx}}$ molecules is
\begin{equation}
    \bar h_{ij}[n]=N_{\mathrm{tx}}\bm{o}_{ij}^{\mathsf T}
    \bm{A}_{ij}^{\mathrm{dec}}(\bm{A}_{ij}^{\mathrm{slot}})^n\bm{b}_{ij},
    \qquad n\geq0.
    \label{eq:single_bit_sampled_response}
\end{equation}
The sample $\bar h_{ij}[0]$ is the mean desired contribution of one bit-1, whereas $\{\bar h_{ij}[n]\}_{n\geq1}$ characterizes its network-slot postcursor memory. The transition-matrix eigendecomposition in~\cite{zheng2025DNA-Based} shows that the molecular-channel response admits a modal representation. Accordingly, we approximate the postcursor tail associated with~\cref{eq:physical_link_state} using $R$ nonnegative decay modes,
\begin{equation}
    \bar h_{ij}[n+1]\approx\sum_{r=1}^{R}w_{ij,r}p_{ij,r}^{n},\qquad n\geq0,
    \label{eq:positive_modal_tail}
\end{equation}
where $w_{ij,r}\geq0$ and $0\leq p_{ij,r}<1$ denote the modal weights and decay factors, respectively. These parameters are obtained by bounded weighted nonlinear least squares over $n=0,\ldots,N_h-1$, using $1/(n+1)$ as the squared-residual weight. The approximation accuracy of each link is evaluated over the same calibration window using the normalized root-mean-square error (NRMSE) and the relative $\ell_2$ error. For the NRMSE, the root-mean-square fitting error is normalized by the largest physical-tail sample.

At the network-slot scale, the fitted decay factors and modal weights are collected into $\bm{\Lambda}_{ij}=\operatorname{diag}(p_{ij,1},\ldots,p_{ij,R})$ and $\bm{w}_{ij}=[w_{ij,1},\ldots,w_{ij,R}]^{\top}$, respectively. The reduced residual state then evolves as
\begin{equation}
    \bm{z}_{ij}[t+1]=\bm{\Lambda}_{ij}\bm{z}_{ij}[t]+\bm{w}_{ij}B_p^{(1)}\sigma_{ij}[t].
    \label{eq:reduced_link_state}
\end{equation}
Each attempted packet transmission therefore adds the calibrated tail generated by its $B_p^{(1)}$ bit-1 releases, regardless of the decoding outcome, whereas an idle link undergoes only modal decay.

The residual molecular contribution available at the routing decision in slot $t$ follows directly from the reduced modal state,
\begin{equation}
    \rho_{ij}[t]\triangleq\bm{1}^{\top}\bm{z}_{ij}[t],
    \label{eq:residual_memory}
\end{equation}
with repeated link activations increasing the residual contribution and idle slots allowing the modal state to decay. To quantify the intrinsic memory strength of each physical link, we define the offline memory index
\begin{equation}
    \chi_{ij}\triangleq\frac{\sum_{n=1}^{N_h}\bar h_{ij}[n]}{\sum_{n=0}^{N_h}\bar h_{ij}[n]},
    \label{eq:link_memory_index}
\end{equation}
where $N_h$ denotes the impulse-response horizon used during link calibration. A larger $\chi_{ij}$ indicates that a larger fraction of the molecular response persists beyond the first network-slot observation. Aggregating the link-level indices over the network gives
\begin{equation}
    \overline{\chi}=\frac{1}{|\mathcal{E}|}\sum_{(i,j)\in\mathcal{E}}\chi_{ij},
    \label{eq:network_memory_index}
\end{equation}
with larger $\overline{\chi}$ corresponding to a stronger overall molecular-memory regime.

\subsection{Memory-Dependent Reliability and Cross-Layer State}
\label{subsec:cross_layer_interface}

Given the residual molecular contribution $\rho_{ij}[t]$ before the routing decision in slot $t$, the receiver observations under the bit-0 and bit-1 hypotheses are modeled as
\begin{subequations}
\label{eq:poisson_observation}
\begin{align}
    Y_{ij}^{(0)}[t] &\sim \operatorname{Poisson}\left(\lambda_{ij}^{\mathrm{bg}}+\rho_{ij}[t]\right),
    \label{eq:poisson_observation_0}\\
    Y_{ij}^{(1)}[t] &\sim \operatorname{Poisson}\left(\lambda_{ij}^{\mathrm{bg}}+\rho_{ij}[t]+s_{ij}\right),
    \label{eq:poisson_observation_1}
\end{align}
\end{subequations}
where $\lambda_{ij}^{\mathrm{bg}}$ denotes the background molecular count and $s_{ij}=\bar h_{ij}[0]$ is the calibrated single-bit desired contribution. Each integer threshold $\eta_{ij}$ is fixed before routing evaluation by minimizing isolated-link bit error and is shared by all policies. For the current pre-packet residual, define the false-alarm and miss probabilities
\begin{subequations}
\label{eq:conditional_bit_errors}
\begin{align}
P_{ij}^{\mathrm{FA}}[t]&=\Pr\{Y_{ij}^{(0)}[t]\geq\eta_{ij}\},\\
P_{ij}^{\mathrm{M}}[t]&=\Pr\{Y_{ij}^{(1)}[t]<\eta_{ij}\}.
\end{align}
\end{subequations}
Conditioned on the actual packet composition, the independent-bit packet-success approximation becomes
\begin{equation}
    P_{ij,p}^{\mathrm{s}}[t]
    =\left(1-P_{ij}^{\mathrm{FA}}[t]\right)^{B_{\mathrm{pkt}}-B_p^{(1)}}
    \left(1-P_{ij}^{\mathrm{M}}[t]\right)^{B_p^{(1)}}.
    \label{eq:packet_success_probability}
\end{equation}
This network-slot abstraction uses the common pre-packet residual $\rho_{ij}[t]$ for all bit decisions and aggregates the packet's bit-1 postcursors in~\cref{eq:reduced_link_state}; it does not explicitly resolve intra-packet state evolution. We therefore validate its network-slot residual and packet-success traces against the corresponding full spatial Markov response. Hence, instantaneous reliability depends on both recent link use and the composition of the head-of-line packet.

\begin{figure}[t]
\centering
\includegraphics[width=0.95\columnwidth]{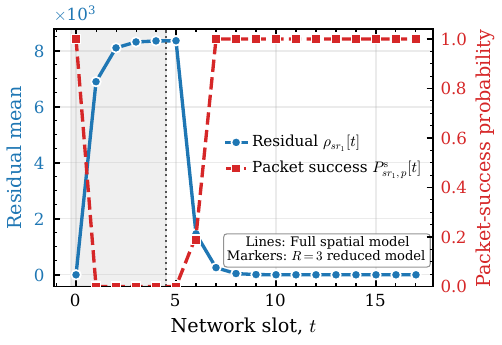}
\caption{Residual-memory accumulation and idle clearing on link $s\!\to\!r_1$. Lines are obtained from the full spatial model in~\cref{eq:physical_link_state}, whereas markers are obtained from the $R=3$ reduced model in~\cref{eq:reduced_link_state}. From a cleared state, one packet containing $B_p^{(1)}=8$ bit-1 symbols is transmitted per slot for $t=0,\ldots,4$, followed by idle slots.}
\label{fig:memory_mechanism}
\vspace{-0.5cm}
\end{figure}

\Cref{fig:memory_mechanism} demonstrates how recent use of the representative slow-clearing link $s\!\to\!r_1$ affects its reliability. During consecutive transmissions, residual molecular contributions accumulate and progressively reduce the composition-conditioned packet-success probability, even though the packet composition and static link parameters remain unchanged. Once transmission ceases, modal decay gradually clears the residual memory and restores the packet-success probability toward its cleared-link value. The reduced-model markers nearly coincide with the corresponding full-spatial curves. For link $s\!\to\!r_1$, the residual trace has a relative $\ell_2$ error of $2.58\times10^{-4}$, while the maximum absolute error in packet-success probability is $1.45\times10^{-2}$. Across all eight directed links, the $R=3$ tail approximations achieve an NRMSE no larger than $3.07\times10^{-4}$. These results confirm that the reduced state preserves the relevant accumulation-and-clearing behavior. Hence, molecular-link reliability depends not only on static physical properties but also on recent transmission history.

To expose this history dependence to the routing layer without propagating the full physical Markov state, each outgoing link provides the quantized cross-layer descriptors
\begin{equation}
c_{ij}^{\rho}[t]=\mathcal{Q}_{\rho}\left(\rho_{ij}[t]\right),\qquad c_{ij,p}^{\mathrm{s}}[t]=\mathcal{Q}_{\mathrm{s}}\left(P_{ij,p}^{\mathrm{s}}[t]\right).
\label{eq:quantized_cross_layer_descriptors}
\end{equation}
where $\mathcal{Q}_{\rho}(\cdot)$ and $\mathcal{Q}_{\mathrm{s}}(\cdot)$ map residual memory and packet-success probability, respectively, into three classes. Accordingly, $c_{ij}^{\rho}[t],c_{ij,p}^{\mathrm{s}}[t]\in\{0,1,2\}$. The two thresholds of each link descriptor are empirical tertiles obtained from an independent nominal-load calibration trace and frozen before Q-training and evaluation. Although $c_{ij,p}^{\mathrm{s}}[t]$ is derived partly from $\rho_{ij}[t]$, it additionally captures packet composition and link-specific signal and threshold parameters. Together, the descriptors provide a compact physical-to-network interface without exposing the spatial Markov state.

\section{Freshness-Aware Routing Formulation}
\label{sec:formulation}
In this section, we formulate the freshness-aware routing problem in terms of packet freshness, delivery deadlines, local observations, and routing actions. For packet $p$, $t_p^{\mathrm{gen}}$ denotes its generation slot and $t_p^{\mathrm{del}}$ its successful delivery slot at the gateway, with $t_p^{\mathrm{del}}=\infty$ for an undelivered packet. Since a transmission outcome is realized at the end of a network slot, the generation time of the freshest update available at the gateway at the beginning of slot $t$ is
\begin{equation}
    U[t]=\max\left(\left\{t_p^{\mathrm{gen}}:t_p^{\mathrm{del}}<t\right\}\cup\{U_0\}\right),
\label{eq:freshest_delivered_update}
\end{equation}
where $U_0\leq0$ denotes the generation time of the update initially available at the gateway. The freshness of the information available at the gateway is measured by the AoI,
\begin{equation}
    \Delta[t]=t-U[t],
    \label{eq:aoi}
\end{equation}
which increases with the time elapsed since the generation of the freshest delivered update. While $\Delta[t]$ characterizes destination-side freshness, routing decisions must act on packets that are still in the network. We therefore use the age of an undelivered packet $p$ at the beginning of slot $t$,
\begin{equation}
    \tau_p[t]\triangleq t-t_p^{\mathrm{gen}},
    \label{eq:packet_age}
\end{equation}
as a local measure of its delivery urgency. Packet delivery is subject to an end-to-end deadline $\tau_{\mathrm{dl}}$, measured in network slots. A packet delivered in slot $t_p^{\mathrm{del}}$ is considered on time if
\begin{equation}
    \tau_p[t_p^{\mathrm{del}}]+1=t_p^{\mathrm{del}}-t_p^{\mathrm{gen}}+1\leq \tau_{\mathrm{dl}}.
    \label{eq:deadline_satisfaction}
\end{equation}
Consequently, an undelivered packet is removed from the network at the beginning of slot $t$ whenever $\tau_p[t]\geq\tau_{\mathrm{dl}}$.

To distinguish a delivery that actually refreshes the gateway from an out-of-order stale delivery, let $\mathcal{D}_g[t]$ be the packets delivered at the end of slot $t$. If this set is nonempty, let $p_t^\star$ be its member with the largest generation time (with sequence identifier breaking ties). We define
\begin{equation}
    \begin{aligned}
        I_p^{\mathrm{info}}[t]
        &=\ind\!\left\{p=p_t^\star,\;t_p^{\mathrm{gen}}>U[t]\right\},\\
        G_p^{\mathrm{AoI}}[t]
        &=I_p^{\mathrm{info}}[t]\left(t_p^{\mathrm{gen}}-U[t]\right),
    \end{aligned}
    \label{eq:informative_delivery_gain}
\end{equation}
where both quantities are zero when $\mathcal{D}_g[t]=\varnothing$. Thus, only the freshest simultaneous delivery can advance the gateway timestamp, $U[t+1]=\max\{U[t],\max_{p\in\mathcal{D}_g[t]}t_p^{\mathrm{gen}}\}$, and $G_p^{\mathrm{AoI}}[t]$ is the resulting one-slot reduction relative to receiving no update. This convention prevents stale or dominated simultaneous packets from receiving freshness credit.

To evaluate the resulting routing performance over a measurement interval of $T$ slots, we first compute the average gateway AoI as
\begin{equation}
    \overline{\Delta}=\frac{1}{T}\sum_{t=1}^{T}\Delta[t],
    \label{eq:average_aoi}
\end{equation}
which captures the long-term freshness of the status information available at the gateway. Timeliness is further quantified by the fraction of generated packets that reach the gateway within $\tau_{\mathrm{dl}}$ and advance its timestamp. Let $\mathcal{P}_{\mathrm{gen}}$ denote the set of packets generated during the measurement interval and $\mathcal{P}_{\mathrm{on}}^{\mathrm{info}}\subseteq\mathcal{P}_{\mathrm{gen}}$ the subset whose delivery is both on time and informative according to~\cref{eq:informative_delivery_gain}. The informative deadline-delivery ratio is
\begin{equation}
    P_{\mathrm{D}}^{\mathrm{info}}=\frac{|\mathcal{P}_{\mathrm{on}}^{\mathrm{info}}|}{|\mathcal{P}_{\mathrm{gen}}|},
    \label{eq:deadline_delivery_ratio}
\end{equation}
which complements $\overline{\Delta}$ without treating an out-of-order stale arrival as a useful status update.

Since fresh and timely delivery may require different numbers of transmission attempts, we additionally measure the associated molecular expenditure. Let $\mathcal{A}_{T}$ denote the set of transmission attempts during the same measurement interval and $p(\nu)$ the packet associated with attempt $\nu\in\mathcal{A}_{T}$. The average number of molecules consumed per informative on-time delivery is
\begin{equation}
    \overline{M}_{\mathrm{on}}^{\mathrm{info}}=\frac{\displaystyle\sum_{\nu\in\mathcal{A}_{T}}M_{p(\nu)}^{\mathrm{tx}}}{|\mathcal{P}_{\mathrm{on}}^{\mathrm{info}}|},
    \label{eq:molecules_per_ontime_delivery}
\end{equation}
which accounts for all attempted transmissions, including unsuccessful attempts and transmissions of packets that subsequently expire or are dropped.

We next specify the local information available for routing decisions. Let $L_i[t]$ denote the queue occupancy of node $i$. For the head-of-line packet $p$, node $i$ constructs the quantized local descriptors
$c_p^A[t]=\mathcal{Q}_A(\tau_p[t])$,
$c_i^q[t]=\mathcal{Q}_q(L_i[t])$, and
$c_i^E[t]=\mathcal{Q}_E(\widetilde E_i[t]/E_{\max})$,
where $\mathcal{Q}_A(\cdot)$, $\mathcal{Q}_q(\cdot)$, and $\mathcal{Q}_E(\cdot)$ map the packet age, queue occupancy, and normalized molecule reserve, respectively, into discrete routing-state classes. Combining these local descriptors with the link-level memory and reliability descriptors in~\cref{eq:quantized_cross_layer_descriptors} yields the MAQR observation at node $i$,
\begin{equation}
    \bm{o}_{i,p}[t]=\left(i,c_p^A[t],c_i^q[t],c_i^E[t],\left\{c_{ij}^{\rho}[t],c_{ij,p}^{\mathrm{s}}[t]\right\}_{j\in\mathcal{N}_i^{+}}\right),
    \label{eq:local_routing_observation}
\end{equation}
where the node identifier $i$ distinguishes node-specific states within the shared routing policy. The resulting observation jointly captures packet urgency, local congestion, molecule availability, and the memory-dependent reliability of each candidate outgoing link.

Whenever node $i$ has a packet and sufficient molecule reserve for transmission, this observation is mapped to a next-hop action,
\begin{equation}
    a_i[t]=\pi\left(\bm{o}_{i,p}[t]\right)\in\mathcal{N}_i^{+},
    \label{eq:routing_policy}
\end{equation}
which activates the selected outgoing link according to
\begin{equation}
    \sigma_{ij}[t]=\ind\{a_i[t]=j\}.
    \label{eq:routing_action_activation}
\end{equation}
If the queue is empty or the molecule reserve is insufficient, no outgoing link is activated. Any attempted transmission consumes molecules and updates the residual-memory state of the selected link, regardless of decoding success.

\section{Memory-Aware Q-Routing}
\label{sec:maqr}

\subsection{Offline Q-Learning}
\label{subsec:offline_learning_reward}

Classical Q-routing uses local Q-values to select the next hop toward a destination~\cite{boyan1993packet}. We retain this lightweight tabular structure while augmenting the routing observation with the molecular-memory descriptors in~\cref{eq:local_routing_observation}. Since the node identifier is included in $\bm{o}_{i,p}[t]$, a shared Q-function can represent node-specific routing decisions. MAQR is trained offline using the reduced molecular-link model, avoiding molecule-expensive exploration and online Q-value updates during deployment.

FIFO forwarding requires the continuation state to belong to the same packet. For an action taken for packet $p$ at node $i$ in slot $t$, let $d_{i,p}[t]\geq1$ be the number of slots until $p$ next becomes head of line and is eligible for a routing decision, and let $(i_p^+,\bm{o}_{i_p^+,p}^{+})$ denote that node and observation. A failed transmission normally gives $d_{i,p}[t]=1$ and $i_p^+=i$. After a successful nonterminal hop to $j$, the bootstrap is delayed until $p$ becomes head of line at $j$, rather than using the state of a different packet at $t+1$. Gateway delivery, expiry, and buffer dropping are terminal. The event-driven update is
\begin{equation}
    \begin{aligned}
        &Q\!\left(\bm{o}_{i,p}[t],j\right)\leftarrow
        Q\!\left(\bm{o}_{i,p}[t],j\right)+\alpha_t\Big[\mathcal{R}_{i,p}[t,d_{i,p}]\\
        &\quad+\gamma^{d_{i,p}}J_{i,p}[t]
        \max_{j'\in\mathcal{N}_{i_p^+}^{+}}
        Q\!\left(\bm{o}_{i_p^+,p}^{+},j'\right)
        -Q\!\left(\bm{o}_{i,p}[t],j\right)\Big],
    \end{aligned}
    \label{eq:maqr_update}
\end{equation}
where $\alpha_t\in(0,1]$, $\gamma\in[0,1)$, and $J_{i,p}[t]$ is one only when a subsequent decision for $p$ exists. The transition return $\mathcal{R}_{i,p}[t,d_{i,p}]$ contains the transmission reward below and any discounted expiry or dropping penalty incurred before the next decision. No other packet's reward or observation is used in this update.

For cost normalization only, define the bookkeeping reference $M_{\mathrm{ref}}^{\mathrm{cost}}=(B_{\mathrm{pkt}}/2)N_{\mathrm{tx}}$. It is not used to scale the physical response in~\cref{eq:single_bit_sampled_response}. Then
\begin{equation}
    \widetilde M_p^{\mathrm{tx}}\triangleq\frac{M_p^{\mathrm{tx}}}{M_{\mathrm{ref}}^{\mathrm{cost}}}=\frac{2B_p^{(1)}}{B_{\mathrm{pkt}}}.
    \label{eq:normalized_attempt_molecules}
\end{equation}

For packet $p$ transmitted by node $i$ in slot $t$, let $F_i[t]$ indicate transmission failure and $X_i[t]$ indicate packet termination due to expiry or buffer dropping. The immediate packet-specific reward is
\begin{equation}
    \begin{aligned}
        R_{i,p}[t]={}&c_g\frac{G_p^{\mathrm{AoI}}[t]}{\tau_{\mathrm{dl}}}-c_h-c_M\widetilde M_p^{\mathrm{tx}}\\
        &-c_q[L_i[t]-1]^+-c_fF_i[t]-c_eX_i[t].
    \end{aligned}
    \label{eq:maqr_reward}
\end{equation}
The first term rewards only informative deliveries that advance $U[t]$ and scales with the resulting AoI reduction, while the remaining terms account for hop usage, normalized molecule expenditure, queued packets, transmission failure, and packet termination. Packet age is retained in the observation so that urgency influences forwarding before delivery. Because decisions rely on local rather than complete network information,~\cref{eq:maqr_update} is interpreted as a lightweight, partially observable, event-driven Q-routing heuristic for decentralized operation, whose effectiveness is assessed through paired empirical evaluation.

During offline training, an $\varepsilon$-greedy policy balances exploration and exploitation, with both $\varepsilon$ and $\alpha_t$ decreasing linearly across training episodes. The traffic load is drawn from a predefined training set, and the initial link-memory states are randomized to expose the learner to both cleared and memory-loaded conditions. During deployment, exploration is disabled, and the next hop is selected as
\begin{equation}
    a_i^\star[t]=\arg\max_{j\in\mathcal{N}_i^{+}}Q\left(\bm{o}_{i,p}[t],j\right).
    \label{eq:maqr_policy}
\end{equation}
For an observation not encountered during offline training, the policy falls back to the instantaneous memory-aware routing (IMR) rule described next.

\subsection{Baseline Routing Policies}
\label{subsec:routing_baselines}

For comparison, we consider three reference routing policies in addition to MAQR. \emph{Shortest-hop routing (SHR)} selects a fixed next hop that minimizes the remaining hop count, using cleared-link reliability only to break ties. The non-learning \emph{instantaneous memory-aware routing (IMR)} baseline recomputes a dynamic negative-log path cost from the same composition-conditioned link reliability available to MAQR:
\begin{equation}
\begin{aligned}
C_{i,p}^{\mathrm{IMR}}[t]
&=\min_{j\in\mathcal{N}_i^{+}}\big[-\log P_{ij,p}^{\mathrm{s}}[t]+\kappa_h\\
&\qquad+\kappa_M\widetilde M_p^{\mathrm{tx}}+C_{j,p}^{\mathrm{IMR}}[t]\big],\\
C_{g,p}^{\mathrm{IMR}}[t]&=0,
\end{aligned}
\label{eq:instantaneous_memory_routing}
\end{equation}
where $\kappa_h$ and $\kappa_M$ penalize an extra hop and its molecule cost. The DAG costs are evaluated in reverse topological order. IMR is reevaluated after every failure and is memory aware, but it does not use packet age, queue occupancy, molecule reserve, or learning. It therefore tests whether Q-routing adds value beyond a simple instantaneous-reliability heuristic.

The third reference policy, \emph{memory-blind Q-routing (MBQR)}, follows the same event-driven update, reward, training seeds, and local nonlink observation as MAQR, but replaces the instantaneous link descriptors with the quantized average reliability of each outgoing link measured on an independent nominal-load calibration trace. In contrast, MAQR observes both $c_{ij}^{\rho}[t]$ and $c_{ij,p}^{\mathrm{s}}[t]$. Thus, the paired MAQR--MBQR comparison changes only access to instantaneous link-memory information.

\subsection{Online Implementation Complexity}
\label{subsec:implementation_complexity}

During deployment, the computational cost of MAQR is dominated by reduced-state tracking and Q-table lookup. For node $i$ with outgoing degree $d_i=|\mathcal{N}_i^{+}|$, updating the $R$ modal states of all candidate outgoing links requires $O(Rd_i)$ operations per network slot, followed by $O(d_i)$ Q-value comparisons for next-hop selection. Since $R=3$ is fixed, the online computational cost scales linearly with the local node degree and remains independent of the dimension of the underlying spatial Markov model in~\cref{eq:physical_link_state}.

The size of the quantized observation space depends on the numbers of packet-age, queue, reserve, residual-memory, and reliability classes, denoted by $C_A$, $C_q$, $C_E$, $C_\rho$, and $C_{\mathrm{s}}$, respectively. For node $i$, the maximum number of possible MAQR observations is
\begin{equation}
    S_i^{\max}=C_A C_q C_E(C_\rho C_{\mathrm{s}})^{d_i},
    \label{eq:maximum_observation_count}
\end{equation}
where $(C_\rho C_{\mathrm{s}})^{d_i}$ accounts for the joint memory and reliability classes of all candidate outgoing links. Because the node identifier is included in $\bm{o}_{i,p}[t]$, observations from different routing nodes remain distinguishable within the shared Q-function. In practice, only state-action pairs encountered during offline training need to be stored. Across the five training pairs, the sparse MBQR and MAQR tables contain $95$--$108$ and $353$--$394$ visited entries, respectively, with means of $100.8$ and $378.6$. Hence, online deployment requires only a sparse Q-table and an $R$-dimensional modal state for each outgoing link, without propagating the full spatial Markov state or performing neural-network inference.

\section{Numerical Results}
\label{sec:results}

\begin{table}[t]
\centering
\caption{Main physical, traffic, learning, and evaluation parameters.}
\label{tab:parameters}
\scriptsize
\setlength{\tabcolsep}{1.6pt}
\resizebox{\columnwidth}{!}{%
\begin{tabular}{@{}lll@{}}
\toprule
Category & Parameter & Value \\
\midrule
Channel
& $D,\Delta x,\Delta t$ & $150~\mu\mathrm{m}^2/\mathrm{s},\,1~\mu\mathrm{m},\,2.5~\mathrm{ms}$ \\
& $k_{\mathrm{on}},c_p,k_{\mathrm{off}}$ & $6{\times}10^8~\mathrm{M}^{-1}\mathrm{s}^{-1},\,10~\mathrm{nM},\,3~\mathrm{s}^{-1}$ \\
& Link distance $d_{ij}$ & $14$--$29~\mu\mathrm{m}$ \\
& Link flow $v_{ij}$ & $6,\;16~\mu\mathrm{m}/\mathrm{s}$ \\
& Absorption buffer & $5~\mu\mathrm{m}$ \\
& Network slot duration $T_s$ & $2~\mathrm{s}$ \\
& Bit interval $T_b$ & $0.125~\mathrm{s}$ \\
& Background mean $\lambda^{\mathrm{bg}}$ & $4060$ \\
& Threshold $\eta_{ij}$ & Linkwise isolated-bit optimum \\
& Modal order / fit horizon $(R,N_h)$ & $(3,15)$ \\
\midrule
Traffic
& Packet length $B_{\mathrm{pkt}}$ & $16$ bits \\
& Molecules per bit-$1$ $N_{\mathrm{tx}}$ & $10^5$ \\
& End-to-end deadline $\tau_{\mathrm{dl}}$ & $10$ slots \\
& Router queue capacity $Q_{\max}$ & $4$ packets \\
& Reserve capacity $E_{\max}$ & $8\times10^6$ molecules \\
& Recovery $E_{\mathrm{rec}}$ & $6\times10^5$ molecules/slot \\
& Evaluation rates $\lambda$ & $\{0.05,0.10,\ldots,0.30\}$ packets/slot \\
\midrule
Learning
& Training rates $\lambda_{\mathrm{train}}$ & $\{0.08,0.14,0.20,0.26,0.30\}$ packets/slot \\
& $\mathcal{Q}_A/\mathcal{Q}_q$ thresholds & $(3,7)$ slots$/\,(1,3)$ packets \\
& $\mathcal{Q}_E$ thresholds & $(0.25,0.65)$ \\
& $\mathcal{Q}_{\rho},\mathcal{Q}_{\mathrm{s}}$ thresholds & Linkwise nominal tertiles \\
& Discount factor $\gamma$ & $0.95$ \\
& Learning rate $\alpha$ & $0.50\rightarrow0.05$ \\
& Exploration rate $\varepsilon$ & $1.00\rightarrow0.02$ \\
& $(c_g,c_h,c_M,c_q,c_f,c_e)$ & $(18,1,0.25,0.20,3,6)$ \\
& IMR $(\kappa_h,\kappa_M)$ & $(1,0.25)$ \\
& Training episodes / length & $5000$ episodes$/\,220$ slots \\
& Independent training seeds & $5$ \\
\midrule
Evaluation
& Warm-up / measured duration & $10^4/10^5$ slots \\
& Paired evaluation seeds & $20$ \\
\bottomrule
\end{tabular}
}
\vspace{-0.5cm}
\end{table}

\begin{figure*}[t]
\centering
\subfigure[Average gateway AoI]{%
\begin{minipage}{0.32\textwidth}
\centering
\includegraphics[width=\linewidth]{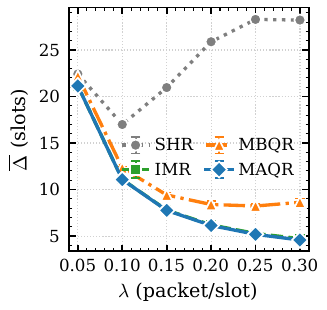}
\end{minipage}
\label{fig:traffic_load_aoi}}
\hfill
\subfigure[Informative deadline-delivery ratio]{%
\begin{minipage}{0.32\textwidth}
\centering
\includegraphics[width=\linewidth]{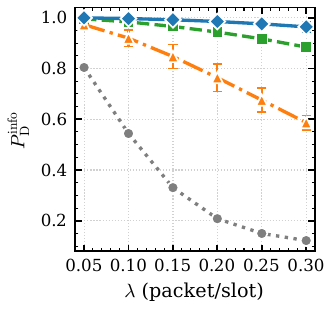}
\end{minipage}
\label{fig:traffic_load_deadline}}
\hfill
\subfigure[Molecules per informative deadline delivery]{%
\begin{minipage}{0.32\textwidth}
\centering
\includegraphics[width=\linewidth]{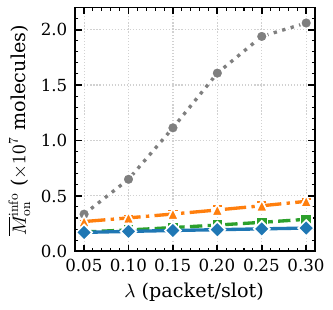}
\end{minipage}
\label{fig:traffic_load_molecules}}
\caption{Routing performance of SHR, IMR, MBQR, and MAQR versus update-generation rate $\lambda$. Error bars indicate hierarchical 95\% bootstrap confidence intervals.}
\label{fig:traffic_load_results}
\vspace{-0.5cm}
\end{figure*}

\subsection{Simulation Setup}
\label{subsec:simulation_setup}

We consider the six-node topology in~\cref{fig:system_model}, with link distances ranging from $14$ to $29~\mu\mathrm{m}$. Links $(s,r_1)$ and $(r_1,g)$ use a flow velocity of $6~\mu\mathrm{m}/\mathrm{s}$, whereas the remaining directed links use $16~\mu\mathrm{m}/\mathrm{s}$, with flow aligned toward the receiving node. This creates a routing tradeoff between the shorter but memory-sensitive route $s\!\to\!r_1\!\to\!g$ and longer, faster-clearing alternatives. Routing uses the $R=3$ reduced model in~\cref{eq:reduced_link_state}, while the full spatial model in~\cref{eq:physical_link_state} is used for offline calibration and the validation in~\cref{fig:memory_mechanism}. Unless otherwise stated, the parameters in~\cref{tab:parameters} are used in our custom Python simulator.

During offline training, each episode samples $\lambda_{\mathrm{train}}\in\{0.08,0.14,0.20,0.26,0.30\}$ and randomized initial residual states. Five independent MAQR--MBQR training-seed pairs share scenario randomness but use independent exploration streams. Descriptor thresholds and MBQR average reliabilities are obtained from a separate nominal-condition calibration trace. The resulting policies are frozen across all traffic-load and physical-memory settings. Each trained pair is evaluated over $10^5$ measured slots after a $10^4$-slot warm-up using 20 common evaluation seeds and common random numbers for packet arrivals, compositions, and link outcomes. Error bars show hierarchical 95\% bootstrap confidence intervals over training pairs and evaluation seeds, whereas statistical comparisons use paired policy differences.

\subsection{Routing Performance versus Traffic Load}
\label{subsec:traffic_results}

\Cref{fig:traffic_load_results} compares SHR, IMR, MBQR, and MAQR for update-generation rates $\lambda=0.05$--$0.30$ packets/slot using the metrics defined in~\cref{eq:average_aoi,eq:deadline_delivery_ratio,eq:molecules_per_ontime_delivery}.
\Cref{fig:traffic_load_aoi} shows the average AoI. As $\lambda$ increases, MAQR and IMR benefit from more frequent status updates and maintain a decreasing AoI trend. In contrast, the improvement of MBQR saturates under higher traffic, while SHR deteriorates beyond $\lambda=0.10$, indicating that its fixed short route cannot accommodate the increasing traffic and residual-memory accumulation. At $\lambda=0.30$, MAQR achieves an average AoI of $4.58$ slots, compared with $8.63$ slots for MBQR and $4.73$ slots for IMR. This corresponds to a $46.9\%$ reduction relative to MBQR, while MAQR remains $0.149$ slots lower than the instantaneous-reliability baseline.

\Cref{fig:traffic_load_deadline} reports the informative deadline-delivery ratio. MAQR maintains $P_{\mathrm{D}}^{\mathrm{info}}\geq0.964$ throughout the load sweep, demonstrating that most generated updates remain both timely and informative even under increasing traffic. By comparison, the ratios of MBQR and SHR decrease substantially with load. At $\lambda=0.30$, MAQR attains $P_{\mathrm{D}}^{\mathrm{info}}=0.964$, whereas MBQR and IMR attain $0.586$ and $0.884$, respectively. The increasing separation from MBQR shows that instantaneous memory descriptors become more important as transmissions occur more frequently.

\Cref{fig:traffic_load_molecules} evaluates the molecular expenditure per informative deadline delivery. MAQR achieves the lowest expenditure across all evaluated loads, showing that its AoI improvement is not obtained by indiscriminately increasing molecule consumption. At $\lambda=0.30$, MAQR requires $2.09\times10^6$ molecules per informative deadline delivery, compared with $4.50\times10^6$ for MBQR and $2.88\times10^6$ for IMR. MAQR therefore reduces molecular expenditure by $53.5\%$ relative to MBQR and by $0.789\times10^6$ molecules relative to IMR.

Paired 95\% bootstrap confidence intervals favor MAQR over MBQR and IMR across all evaluated loads and metrics. The unseen-observation fallback fraction remains below $2.8\times10^{-4}$ of learned-policy transmission attempts. At $\lambda=0.20$, MAQR selects the short source link with conditional probabilities $1.00$ and $0.00$ in the low- and high-memory classes, respectively, confirming memory-dependent route switching.

\subsection{Frozen-Policy Performance versus Physical Memory}
\label{subsec:memory_results}

To vary the physical channel memory in a controlled manner, we introduce a memory scale $\zeta$ and scale the molecular unbinding rate and link flow velocity as
\begin{equation}
    k_{\mathrm{off}}(\zeta)=\frac{k_{\mathrm{off}}}{\zeta},\qquad
    v_{ij}(\zeta)=\frac{v_{ij}}{\sqrt{\zeta}}.
    \label{eq:physical_memory_family}
\end{equation}
Reversible unbinding and advective flow are varied because they govern molecular retention and clearance in the underlying Markov channel~\cite{zheng2026system}. To isolate tail persistence from cleared-link strength, we normalize each response before fitting the reduced modes:
\begin{equation}
    \bar h_{ij}^{\mathrm{sweep}}[n;\zeta]
    =
    \bar h_{ij}[n;\zeta]
    \frac{\bar h_{ij}[0;1]}{\bar h_{ij}[0;\zeta]}.
    \label{eq:memory_sweep_normalization}
\end{equation}
This normalization fixes the single-bit sample at its nominal value while preserving the $\zeta$-dependent tail shape and is applied only in the controlled memory sweep. We evaluate $\zeta\in\{0.20,0.45,0.70,1.00,1.25\}$ at $\lambda=0.20$ packets/slot and recompute $\bar{\chi}$ from~\cref{eq:link_memory_index}.

\begin{figure}[t]
\centering
\includegraphics[width=0.9\columnwidth]{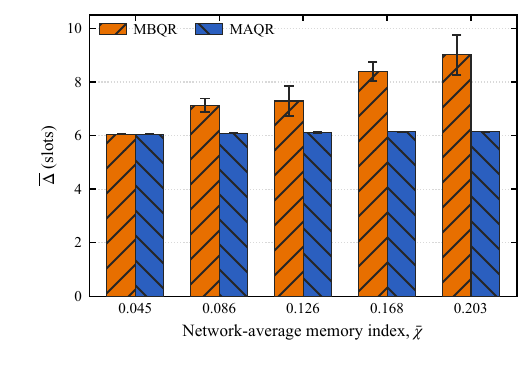}
\caption{Average gateway AoI of frozen MBQR and MAQR versus network-average memory index $\bar{\chi}$ at $\lambda=0.20$ packets/slot. Error bars indicate hierarchical 95\% bootstrap confidence intervals.}
\label{fig:memory_performance}
\vspace{-0.5cm}
\end{figure}

\Cref{fig:memory_performance} shows that MBQR and MAQR perform comparably under weak physical memory but separate as memory persistence increases. At the weakest-memory point, $\bar{\chi}=0.045$, MBQR and MAQR attain $6.049$ and $6.054$ slots, respectively, and their paired 95\% confidence interval includes zero. The paired intervals at the remaining four memory points exclude zero in the direction favoring MAQR. At the strongest-memory point, $\bar{\chi}=0.203$, MBQR and MAQR attain $9.018$ and $6.145$ slots, respectively, corresponding to a $2.874$-slot, or $31.9\%$, AoI reduction.

The MBQR unseen-observation fraction reaches $91.1\%$ at the weakest-memory point, falls below $0.001\%$ at the nominal condition, and rises to $18.2\%$ at the strongest-memory point, whereas the MAQR fraction remains below $0.12\%$ throughout. Because both agents invoke the same dynamic IMR rule for unseen observations, the figure represents complete frozen-policy deployment performance, including state-coverage and fallback effects, rather than a descriptor-only causal comparison.

\section{Conclusion}

This paper developed MAQR as a lightweight, partially observable Q-routing heuristic for fresh status delivery over history-dependent molecular links. Its single-bit calibration and composition-conditioned reliability connect physical link memory with routing decisions, while event-driven FIFO updates and an AoI-reduction reward align learning with information freshness. At the highest evaluated load of $0.30$ packets/slot, MAQR reduces average AoI by $46.9\%$ relative to MBQR while also improving the informative deadline-delivery ratio and molecular efficiency. Smaller but consistent improvements are observed over IMR. The frozen-policy memory sweep further shows comparable performance under weak memory and an increasing MAQR advantage as physical memory strengthens. These results demonstrate the value of reduced link-memory states for adaptive molecular routing.

Future work will validate frozen policies using the full spatial model and experimentally calibrated molecular channels, and will quantify robustness to model order, physical-model mismatch, reward design, and descriptor selection. Scalable function approximation, multi-agent learning, and online adaptation to time-varying traffic and channel conditions will also be investigated for larger multi-source IoBNT networks.
\bibliographystyle{IEEEtran}
\bibliography{refs}

\end{document}